\pdfoutput=1
\RequirePackage{fix-cm}
\documentclass[british]{article}

\usepackage{spconf} 

\usepackage[utf8]{inputenc}
\usepackage[T1]{fontenc}
\usepackage[british]{babel}
\usepackage{float}

\usepackage{pifont}

\usepackage[pdftex]{graphicx}
\usepackage[table,xcdraw]{xcolor}
\graphicspath{{./figures/}}
\DeclareGraphicsExtensions{.pdf,.png,.jpeg,.jpg}

\usepackage[cmex10]{amsmath} 
\usepackage{amssymb}
\usepackage{bm} 
\usepackage{esint} 
\usepackage{commath} 
\usepackage{units} 

\usepackage{subcaption} 

\usepackage{booktabs} 
\usepackage{array} 

\usepackage{multirow} 
\usepackage{rotating} 

\usepackage{enumitem} 
\setlist{nolistsep} 

\usepackage[unicode=true,pdfusetitle,%
 pdfauthor={Gustavo Teodoro Döhler Beck, Ulme Wennberg, Zofia Malisz, Gustav Eje Henter},%
 pdfkeywords={speech synthesis control, speech science methods, neural vocoder, adversarial learning, speech perception},%
 bookmarks=true,bookmarksnumbered=true,bookmarksopen=false,%
 breaklinks=true,pdfborder={0 0 0},backref=false,colorlinks=true,citecolor=blue]{hyperref}
\usepackage[capitalise]{cleveref} 

\renewcommand{\mid}{\,\ifnum\currentgrouptype=16 \middle\fi|\,}

\ninept 

\let\oldmarginpar\marginpar
\renewcommand\marginpar[1]{\-\oldmarginpar[\raggedleft\footnotesize #1]%
{\raggedright\footnotesize #1}}

\title{Wavebender GAN:\\{}An architecture for phonetically meaningful speech manipulation}

\name{Gustavo Teodoro Döhler Beck, Ulme Wennberg, Zofia Malisz, Gustav Eje Henter\thanks{This work was partially supported by the Wallenberg AI, Autonomous Systems and Software Program (WASP) funded by the Knut and Alice Wallenberg Foundation. ZM was supported by the Swedish Research Council grant no.\ 2017-02861 ``Multimodal encoding of prosodic prominence in voiced and whispered speech''.}}
\address{Division of Speech, Music and Hearing, KTH Royal Institute of Technology, Stockholm, Sweden}

\begin{document}
\maketitle
\begin{abstract}
Deep learning has revolutionised synthetic speech quality. However, it has thus far delivered little value to the speech science community. The new methods do not meet the controllability demands that practitioners in this area require e.g.: in listening tests with manipulated speech stimuli.
Instead, control of different speech properties in such stimuli is achieved by using legacy signal-processing methods. This limits the range, accuracy, and speech quality of the manipulations.
Also, audible artefacts have a negative impact on the methodological validity of results in speech perception studies.

This work introduces a system capable of
manipulating speech properties
through learning rather than design. The architecture learns to control arbitrary speech properties and leverages progress in neural vocoders to obtain realistic output.
Experiments with copy synthesis and manipulation of a small set of core speech features (pitch, formants, and voice quality measures) illustrate the promise of the approach for producing speech stimuli that have accurate control and high perceptual quality.
\end{abstract}
\section{Introduction}
\label{sec:intro}
Speech synthesis based on data-driven machine learning, e.g., WaveNet \cite{vandenoord2016wavenet}, has been shown to exceed previous realism standards with a margin \cite{shen2018natural}. These innovations, however, have not yet been taken up into the toolbox of methods employed in basic speech and language research. The reason is that modern synthesis systems do not offer adequate control over low-level acoustic parameters, such as duration, formant frequencies, or fundamental frequency.

Phonetic and psycholinguistic experiments that aim to disentangle the perceptual role of particular acoustic features in the speech signal often involve the creation of well-controlled audio stimuli using speech synthesis. To this day, these stimuli are created by legacy systems, e.g.: formant synthesis (e.g., the Klatt synthesiser \cite{klatt1990analysis} or OVE \cite{carlson1982multi}) or overlap-add techniques, e.g., \cite{charpentier1986diphone}.
%
%
However, speech generated by legacy systems is processed differently by humans than natural speech. It overburdens attention and cognitive mechanisms, evident through slower processing times \cite{winters2004perception}. The signal does not preserve the natural variability or acoustic-phonetic cue redundancies that are fundamental in human perception of natural speech. This casts doubt on the universality of research findings associated with using synthetic stimuli generated by legacy systems \cite{winters2004perception,van1999categorical}.

In this project, we aim to create a synthesis method that combines the precise control over phonetically interpretable acoustic parameters, such as duration, formant frequencies, or pitch, provided by signal-processing methods, with the perceptual closeness to natural human speech offered by modern deep-learning techniques.

We introduce a \emph{general} architecture designed to manipulate perceptually relevant features in a speech-in-speech-out system that leverages a pre-trained neural vocoder. This means the model we propose is not contingent on manipulating one, or a specific set, of features. In practice, we note, model performance will depend on what feature set is used, as it is important for features to be both a) phonetically meaningful and b) sufficiently disentangled to achieve speech reconstruction and manipulation, respectively. We \emph{validate} the approach in terms signal reconstruction quality and manipulation effectiveness on a specific set of core features from phonetics.
Audio examples are available at \href{https://gustavo-beck.github.io/wavebender-gan/}{gustavo-beck.github.io/wavebender-gan/}.





\section{Prior work}
\label{sec:background}


Speech manipulation has a long history as part of the bread and butter of phonetic research.
Classic formant synthesis methods \cite{klatt1990analysis,carlson1982multi} can produce high-quality, manipulated synthetic speech \cite{hunt1989issues}.
However, the process is laborious and most stimuli sound distinctly unnatural \cite{winters2004perception,malisz2019modern}. 
To manipulate pitch and duration, many signal-processing-based techniques exist, e.g.:  pitch-synchronous overlap-add (PSOLA) \cite{charpentier1986diphone}.
These operate in the waveform-domain and offer good quality with noticeable perceptual artefacts. Signal processing-based vocoders such as STRAIGHT \cite{kawahara2006straight} allow explicit pitch manipulation and can be used in speech perception research involving voice morphing \cite{kawahara2015temporally}.
However, the manipulation of other phonetically meaningful properties is less well-served by the signal processing.
For any given perceptual property of interest, manipulation methods may be poorly developed or even non-existent, as bespoke methods require substantial signal processing research.

\begin{figure*}[!t]
    \centering
    \includegraphics[width=0.75\textwidth]{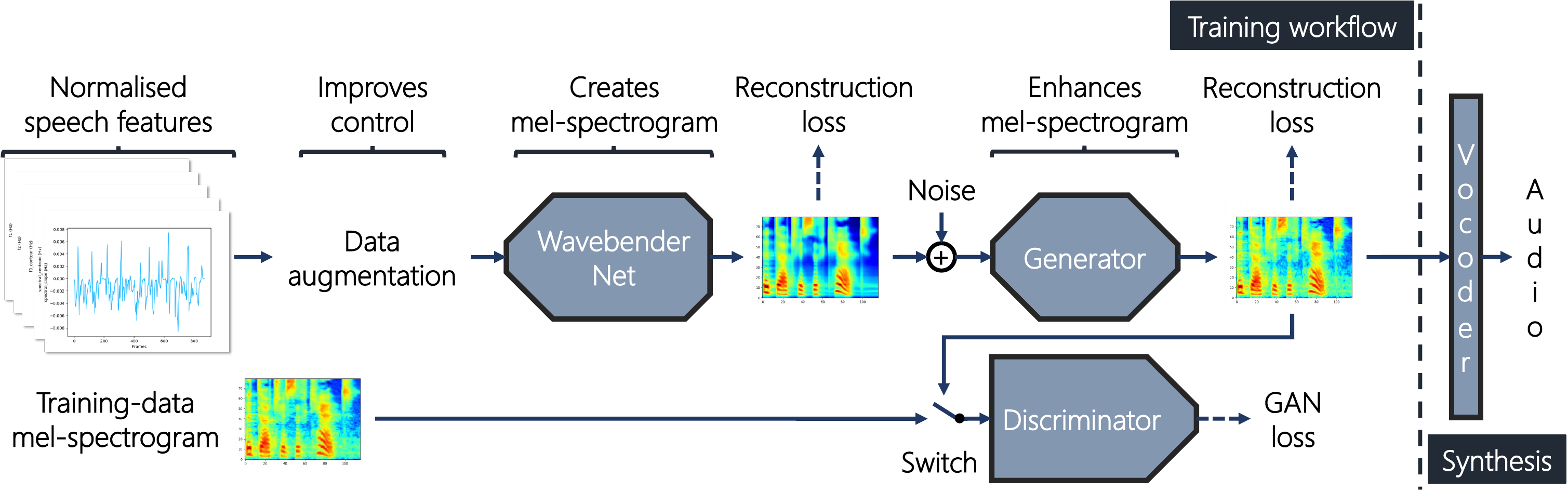}
    \caption{Wavebender GAN workflow for training and synthesis. Network shapes in the figure are meaningful and indicate layer widths.}
    \label{fig:workflow}
    \vspace{-1\baselineskip}
\end{figure*}

Neural vocoders, that generate speech waveforms using deep-learning, have recently emerged as an alternative to classical vocoders such as \cite{kawahara2006straight}.
This enabled a step change in synthetic-speech naturalness, first demonstrated by WaveNet \cite{vandenoord2016wavenet} for TTS and then adapted for neural vocoding in \cite{tamamori2017speaker}.
Unlike classical vocoders based on signal processing, the mapping from input features to output waveform in a neural vocoder is learned from data rather than manually designed.
The same general machine-learning approach can therefore be used to create vocoders with many different control interfaces, and the resulting signal quality is not limited by signal-processing but by the machine-learning methodology and 
data quality.
This article proposes a concrete deep-learning framework for controlling arbitrary speech properties validated to give both high signal quality and accurate control.

To date, many high-quality neural vocoders have been proposed, harnessing a variety of deep-learning approaches including dilated CNNs \cite{vandenoord2016wavenet}, RNNs \cite{mehri2017samplernn,kalchbrenner2018efficient,valin2019lpcnet}, GANs \cite{bollepalli2017generative,juvela2019gelp,kong2020hifi}, normalising flows \cite{prenger2019waveglow,kim2019flowavenet}, and denoising diffusion models \cite{kong2021diffwave,chen2021wavegrad}.
At present, the most widely-used neural vocoder is likely WaveGlow \cite{prenger2019waveglow}, although HiFi-GAN \cite{kong2020hifi} has emerged as a drop-in replacement with state-of-the-art speech signal quality.

Most neural vocoders 
synthesise speech waveforms from (mel) spectrogram acoustic features \cite{tan2021survey}.
While these input features are easy to extract from speech and work well in text-to-speech applications,
they are not an intuitive representation for speech manipulation. Neural vocoders \cite{bollepalli2017generative,valin2019lpcnet,juvela2019gelp} that leverage the source-filter model of speech production \cite{fant1960acoustic} offer a better starting point for pitch manipulation, but not necessarily other relevant
speech properties.
Our deep-learning-based approach to speech manipulation provides simultaneous control over many phonetically meaningful speech properties chosen by the experimenter, for example formant frequencies and voice-quality measures like spectral slope.

A system similar to ours is demonstrated in \cite{juvela2018speech}: a GAN-based model for speaker-specific synthesis of high-quality speech waveforms, specifically, from 20 mel-frequency cepstrum coefficients (MFCCs) per analysis frame. 
Our work improves on \cite{juvela2018speech} by a) avoiding the need to train new models at the waveform level by leveraging existing high-quality vocoders such as \cite{kong2020hifi}, b) achieving realistic synthesis using only 5 speech parameters per frame (plus a binary flag), thus further advancing minimalist vocoder interfaces, and c) parameterising the control in terms of phonetically relevant speech properties rather than MFCCs.

\section{Method}
\label{sec:method}
We describe the proposed methodology for controllable speech synthesis from perceptually-meaningful speech parameters
, which comprises two major steps:
\begin{enumerate}
\item A modified ResNet that generates vocoder features
(here mel-spectrograms) from selected speech parameters (Sec.\ \ref{ssec:resnet})
\item A post-processing network that enhances the output from the previous network, trained using a GAN loss (Sec.\ \ref{ssec:gan})
\end{enumerate}
The final vocoder features (mel-spectrograms) are then fed into a pre-trained neural vocoder to generate natural-sounding synthetic speech waveforms.
The proposed approach is visualised in Fig.\ \ref{fig:workflow}.




\subsection{Problem formulation}
\label{ssec:problem}
Conceptually, our task of synthesising speech from a set of (phonetically relevant) speech parameters can be cast as the inverse of feature extraction.
We assume that feature extractors exist to compute all the relevant features from recorded speech, at a particular framerate.
We apply the feature extractors to a speech dataset to obtain a training set that includes waveforms and associated sequences of speech parameters.
While the performance of our approach is limited by the quality of the feature extractors, improved speech parameter extraction is out of scope of this work.
In general, we expect the accuracy of off-the-shelf feature extractors to keep increasing, bringing steady improvements to our method.



Formally speaking, the task we consider in this work is to take a series of input vectors, representing the speech properties of interest at each time frame, and use these to generate a synthetic speech waveform.
We leverage a (pre-trained neural) vocoder for waveform generation to make use of existing advances in speech signal generation. This way we reduce the original problem to the mapping of input vectors to an output vectors series (mel-spectrograms) that, in turn, constitute input features to the vocoder.
From now on, we assume that the vocoder features are log-energy mel-spectrograms since these are widely used \cite{tan2021survey} and easy to extract from audio.
The resulting setup amounts to a sequence-to-sequence regression problem where the input and output sequences are already time-aligned.

\subsection{Wavebender Net for initial mel-spectrogram creation}
\label{ssec:resnet}
The fact that inputs and output series are aligned in time suggests using a convolutional architecture to map between the two.
We perform the initial mapping from input to output sequence using an architecture inspired by the successful ResNet architecture \cite{he2016deep} from computer vision, which we adapt and enhance for the task at hand.
Our modifications are
1) changing the architecture from 2D to 1D convolutions, since we are working on time series (one dimension) instead of images that have both width and height;
2) adding long-range skip connections which were found to improve learning in U-Net \cite{ronneberger2015u} and FlowNet \cite{dosovitskiy2015flownet}; and
3) replacing batchnorm by group normalisation \cite{wu2018group} which performs better on small batches as considered here.
We call this component \emph{Wavebender Net}.
Like all other networks in the paper, we make this network wider in its intermediate layers,
to not limit its capacity and benefit from the lottery ticket hypothesis \cite{frankle2019lottery} in learning accurate mappings.

We train our model using
a regression loss (the original ResNet was trained as a classifier).
Specifically%
, we chose the XSigmoid loss function,
instead of the more common L2/MSE loss, since XSigmoid has better statistical robustness to outliers (similar to \cite{henter2016robust}) and leads to sharper, less over-smoothed output.
That said, we still find
minor
artefacts in the generated mel-spectrograms (see Fig.\ \ref{fig:learned_mel_representation}), which is why we introduce the enhancement network in the next section.

In initial experiments we noticed that Wavebender Net worked well for copy-synthesis,
but that changing some of our chosen speech parameters, most noticeably pitch, away from the values in the copy synthesis reduced speech quality. It also did not accurately reflect the desired manipulation value.
We hypothesised that this issue was due to a lack of representative data (e.g., high and low pitches) in the training set.
We solved the problem using data augmentation: by randomly manipulating input speech and re-extracting the associated speech parameters,
training examples can be created that cover a wider region of the speech parameter space.
Importantly, the augmentations used need not be related to the speech parameter set used -- our experiments applied utterance-level f0 manipulation and gain modification, even though gain has no effect on our speech parameters.
This teaches the network to be independent of gain.
Informal testing confirmed that data augmentation noticeably improved control accuracy.

\subsection{Conditional GAN for enhancing mel-spectrograms}
\label{ssec:gan}

Deterministic predictors of stochastic signals, such as speech, generally exhibit averaging artefacts such as oversmoothing.
To reinsert perceptual cues lost in the averaging and remove other artefacts of deterministic prediction, we use a second network (the GAN \emph{generator}) that maps from mel-spectra to enhanced mel-spectra, based on conditional GAN (cGAN) techniques \cite{mirza2014conditional}.
Most cGANs also take random noise as input, 
to enable stochastic synthesis.
We accomplish this by adding 
a small amount of noise
to each spectral bin.
The result is passed through the rest of the generator, which is composed of 2D CNN layers.
The discriminator also consists of 2D CNNs.
We call our complete setup \emph{Wavebender GAN}.

We use an LS-GAN loss \cite{mao2017least} to train the generator and the discriminator. We also compute the XSigmoid reconstruction loss of the generated spectrograms before and after the generator. The final objective function is the sum of these two terms and the LS-GAN loss.
The discriminator, the generator, and Wavebender Net are then trained simultaneously, end-to-end.
To accelerate convergence, training is initialised from a pre-trained Wavebender Net.

\section{Experiments}
\label{sec:experiments}


\subsection{Data and setup}
\label{ssec:data}
We performed our experiments on the LJ Speech dataset \cite{ljspeech17} which comprises 13,100 utterances (approximately 24 h) from a single female speaker of US English reading books aloud.
The data was randomly split into a 95\% training and 5\% testing utterance sets.

For the experiments reported here, we considered a core set of phonetically meaningful, frame-wise speech parameters that we identified in the features provided by the Surfboard library \cite{lenain2020surfboard}.
To construct this set, we first manually identified a subset of the Surfboard features believed to be most widely relevant to phonetic research, and then analysed their correlations using Spearman's $\rho$.
Features were then excluded from the set guided by high numbers in the correlation matrix, since input features with high correlations was found to negatively affect manipulation robustness.

By doing so, we arrived at five phonetically meaningful features, namely: F1 (Hz), F2 (Hz), f0 (voiced/unvoiced binary flag and log frequency, linearly interpolated in unvoiced regions), spectral centroid, and spectral slope. These were used in all  experiments.
The formant frequencies did correlate substantially, but retaining both was nonetheless necessary to be able to perform manipulations over the entire vowel space.
The features were $z$-score normalised (i.e., mean 0, st.\ dev.\ 1) for the network training, which also means that MSE reconstruction errors were comparable between features.
Data augmentation during training was performed using Parselmouth \cite{parselmouth} for pitch manipulation and Surfboard for speech gain.


For the networks, we used 8 residual blocks in Wavebender Net, 6 CNN layers in the generator, and 12 CNN layers in the discriminator (with linear output as proposed in LS-GAN \cite{mao2017least}).
Long-range residual connections inspired by \cite{ronneberger2015u,dosovitskiy2015flownet} were inserted into the Wavebender Net from after block 1 to before block 8 and after block 2 to before block 7.
The Wavebender Net was initially trained for 20 epochs, followed by 10 epochs training of the full system.
Training used Adam \cite{kingma2015adam} with a cosine learning-rate scheduler (cf.\ \cite{sohn2020fixmatch}).

In preliminary copy-synthesis experiments, we found HiFi-GAN \cite{kong2020hifi} to provide significantly better reconstruction accuracy for our selected speech features than WaveGlow \cite{prenger2019waveglow} did, in addition to providing higher perceived signal quality, as reported in \cite{kong2020hifi}.
%
%
We therefore used HiFi-GAN for all experiments reported in this paper.

\subsection{Evaluation}
We evaluated the ability of the system to reconstruct low-level speech features, and to robustly and precisely manipulate these speech features, while simultaneously producing high-quality synthesis.
Additional information such as synthesis samples is provided on the paper webpage at \href{https://gustavo-beck.github.io/wavebender-gan/}{gustavo-beck.github.io/wavebender-gan/}.

First, we performed copy-synthesis and measured speech parameter reconstruction accuracy. We used Wavebender GAN to synthesise on the basis of low-level features obtained from a natural speech recording.
We then
extracted the same parameters from the reconstructed speech and compared the analysis-synthesis error for each parameter. 
We measured the error in each $z$-score normalised feature for every time frame, and then averaged this metric across all time frames and samples. Results are shown in Fig.\ \ref{fig:mse}.






Second, we extended the copy-synthesis experiment to evaluate the effect of manipulating individual speech parameters.
We did this by globally scaling each speech parameter with a scaling factor $m \in \{ 0.7, 0.8, 0.9, 1.0, 1.1, 1.2, 1.3\}$, one parameter at a time, while keeping the remaining parameters intact (except when manipulating F1, F2 where we corrected for correlations as explained below).
These scaled feature trajectories were fed into Wavebender GAN, which fed into HiFi-GAN to synthesise manipulated speech.
We then computed the MSE of the desired vs. the realised feature trajectories across all normalised features, to evaluate how well the manipulated speech respected the control. 

Results are illustrated in Fig.\ \ref{fig:manipulation}.
In addition to that figure, which reports the error in aggregate, the paper webpage presents a matrix that quantifies the effect of these manipulations on each individual speech parameter, as a way to study parameter disentanglement.


As explained in Sec.\ \ref{ssec:data}, F1 and F2 are correlated and may never be decoupled completely (e.g., F2 < F1 is impossible by definition).
Several approaches to handling this are possible, but here (including Fig.\ \ref{fig:manipulation})
we allowed the user to manipulate either one of the features (e.g., the F1), and then use a neural-network architecture similar to Wavebender Net to predict the induced manipulation on the other feature (i.e., a corresponding F2 trajectory that respects the correlation between the two features) to arrive at the final feature trajectories for the manipulation experiment.


Third, we complemented our objective evaluation of synthesis accuracy, with and without signal manipulation, with a subjective listening test. We selected 10 natural speech clips at random from the test set.
We then performed a Mean Opinion Score (MOS) test, comparing the listener preferences for speech generated by HiFi-GAN copy-synthesis and Wavebender GAN copy-synthesis (with HiFi-GAN as the vocoder), relative to the natural speech recording as a reference.
In each trial, listeners were presented with the reference first. Next, they were asked to rate one of the system-generated signals presented consecutively in a counterbalanced order ($N$=20 ratings per participant). We did not specify which system generated a given stimulus. The listeners rated signal naturalness on a 5 point scale (1-bad, 2-poor, 3-fair, 4-good, 5-excellent). Twenty nine listeners completed the survey set up on the Phonic crowdsourcing platform ($N$=580 total ratings).
The results are reported in Sec.\ \ref{mos_test_res}.

\section{Results and discussion}
\label{sec:results}

\subsection{Objective evaluation of reconstruction accuracy}
\label{ssec:accuracy}
\begin{figure}[!t]%
    \centering
    \subfloat[Wavebender Net \label{sfig:net}]{{\includegraphics[width=0.31\columnwidth]{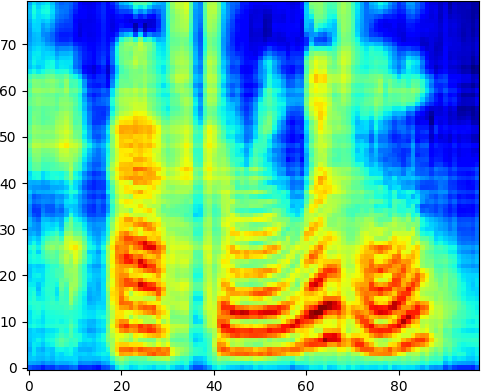} }}%
    \hfill
    \subfloat[Wavebender GAN \label{sfig:gan}]{{\includegraphics[width=0.31\columnwidth]{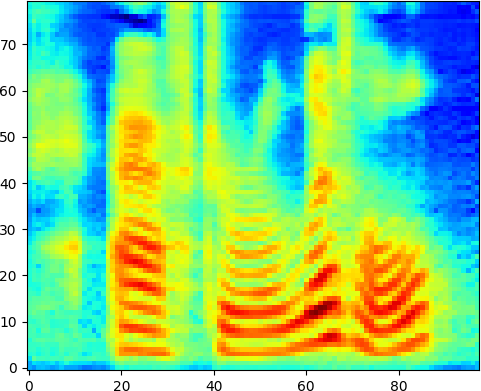} }}%
    \hfill
    \subfloat[Target speech\label{sfig:target}]{{\includegraphics[width=0.31\columnwidth]{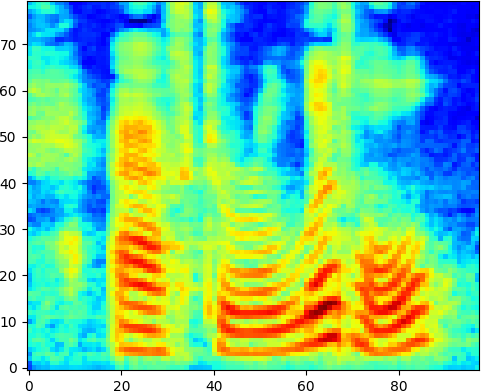} }}%
    \caption{Mel-spectrogram representations generated by Wavebender Net and Wavebender GAN compared to the target representation.}%
    \label{fig:learned_mel_representation}%
    \vspace{-1\baselineskip}
\end{figure}

Fig.\ \ref{fig:learned_mel_representation} compares example mel-spectrograms, in a copy-synthesis setting, produced by Wavebender Net (\ref{sfig:net}) and by the entire Wavebender GAN pipeline (\ref{sfig:gan}) to the mel-spectrogram of target natural speech (\ref{sfig:target}).
We can see Wavebender GAN successfully removes subtle artefacts present in the Wavebender Net output.

\begin{figure}[!t]
    \centering
    \includegraphics[width=\columnwidth]{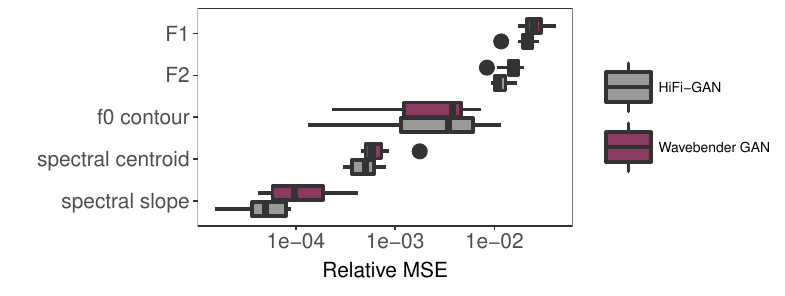}
    \caption{Copy-synthesis error for each speech property for Wavebender GAN$\to$HiFi-GAN vocoder versus HiFi-GAN only.}
    \label{fig:mse}
    \vspace{-1\baselineskip}
\end{figure}

Moreover, the results presented in Fig.\ \ref{fig:mse} indicate that Wavebender GAN adds a relatively minor amount to the reconstruction errors of the different features. All errors in this paper are measured on the normalised features, to make reported values more comparable across features. The majority of the error is attributable to the vocoder (HiFi-GAN) used in the current experiments, suggesting that the vocoder, not Wavebender GAN, could be the main factor limiting control accuracy.
Also, it is apparent that the data augmentation is effective for improving quality, since the Wavebender GAN error for f0 contours is smaller relative to HiFi-GAN. 

\subsection{Objective evaluation of manipulation effectiveness}

\begin{figure}[!t]
    \centering
    \includegraphics[width=\columnwidth]{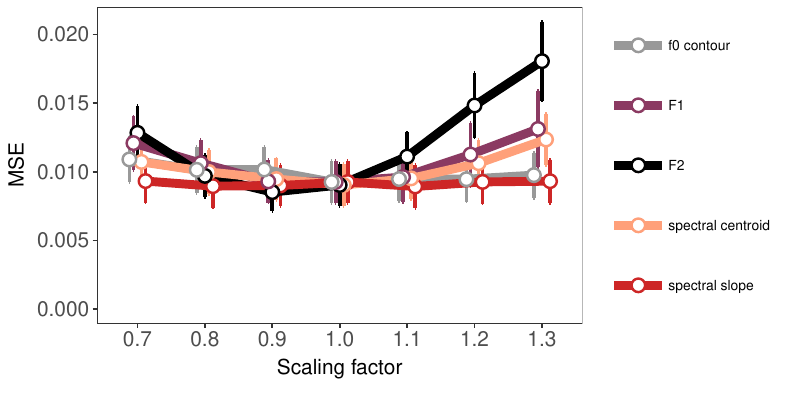}
    \caption{Overall MSE (mean and standard deviation intervals) away from the desired value when using Wavebender GAN to manipulate (scale) each speech feature in isolation.}
    \label{fig:manipulation}
    \vspace{-1\baselineskip}
\end{figure}

Fig.\ \ref{fig:manipulation} shows how the overall error across all features changes as different individual features are manipulated away from pure copy synthesis to perform speech manipulation.
We see that changes to voice-quality features or the f0 contour are well reflected in the output speech and do not inflate the overall error.
Manipulations of F1 and F2, however, are not as accurately reflected in the output features.
We attribute this to the lack of feature augmentation for formant frequencies, in contrast to f0.
A simple alternative to explicit augmentation for improving the coverage of the formant space (and the input space in general) would be to consider a large database of speakers of different genders, accents, and even languages, similar to the development of so-called universal neural vocoders \cite{lorenzo2018towards}.

\subsection{Subjective MOS listening test}
\label{mos_test_res}
First of all, the mean opinion score for the HiFi-GAN copy-synthesis pipeline approximately matches the performance reported in \cite{kong2020hifi} ($4.36\pm 0.07$), despite differences in methodology.
The Wavebender GAN$\to$HiFi-GAN vocoder copy-synthesis pipeline achieves
a mean $4.12\pm 0.31$ rating in the MOS test, which does not substantially deviate from the mean rating obtained by the signals produced by HiFi-GAN copy-synthesis ($4.44\pm 0.16$), and exceeds WaveGlow ($3.81\pm 0.08$) in \cite{kong2020hifi}. 
In addition to speech parameter control, these results mean that we obtained the high output naturalness -- close to a state-of-the-art neural vocoder -- that we had set out achieve.




\section{Conclusions and future work}
\label{sec:conclusion}
We have demonstrated a deep-learning approach to synthesising high-quality speech from a small set of phonetically meaningful speech parameters, as well as the ability to accurately control these parameters in the output speech.
To the best of our knowledge, our pipeline is the first architecture to combine state-of-the-art speech quality with this interpretability and low-level feature control.
Together, these two properties are what is needed to realise a step change in the ability of
speech scientists to construct end-to-end pipelines for stimulus creation and testing of phonological models. 

The main performance limitation appears not to sit in our architecture, but in the neural vocoder.
As neural vocoder technology is advancing rapidly at present, we expect that further gains in quality and accuracy will follow without altering our fundamental approach.

We see several directions for future research, beyond changing the general machine-learning approach, we plan to: i) validate the model on more speakers and other feature sets such as ones that would allow to manipulate voice modes e.g.: whispered or Lombard speech, ii) extend the approach to multiple speakers, both to broaden applicability and to improve control by training on data that better covers all of speech,
and iii) demonstrate the utility of the approach in performing phonetic research.  


\textbf{Acknowledgement:}
The authors thank Jonas Beskow for helpful feedback on the work.


\bibliographystyle{IEEEbib}
\bibliography{refs_abbrev}

\end{document}